\documentclass[sigplan,screen,10pt,natbib=false]{acmart}

\setcopyright{none}
\renewcommand\footnotetextcopyrightpermission[1]{} 
\usepackage[style=numeric]{biblatex}
\usepackage{url}
\usepackage{hyperref}
\bibliography{paper}
\usepackage{mathtools,amsmath,amsfonts,amsxtra,amsthm,mathrsfs,mathpartir}
\usepackage{listings}
\usepackage{xcolor}
\usepackage{array}
\usepackage[toc,page]{appendix}
\usepackage{graphicx}
\usepackage{varioref}
\usepackage{verbatim,comment}
\usepackage{longtable}
\usepackage{tikz}
\usepackage{caption}
\usepackage{subcaption}
\usepackage{microtype}
\usepackage{balance}
\definecolor{OliveGreen}{RGB}{128,128,0}

\usepackage{xurl}
\expandafter\def\expandafter\UrlBreaks\expandafter{\UrlBreaks
  \do\a\do\b\do\c\do\d\do\e\do\f\do\g\do\h\do\i\do\j%
  \do\k\do\l\do\m\do\n\do\o\do\p\do\q\do\r\do\s\do\t%
  \do\u\do\v\do\w\do\x\do\y\do\z\do\A\do\B\do\C\do\D%
  \do\E\do\F\do\G\do\H\do\I\do\J\do\K\do\L\do\M\do\N%
  \do\O\do\P\do\Q\do\R\do\S\do\T\do\U\do\V\do\W\do\X%
  \do\Y\do\Z}

\lstdefinelanguage{nickel}{
    keywords=[1]{
      if,
      then,
      else,
      switch,
    },
    keywords=[2]{
      let,
      rec,
      fun,
      in
    },
    keywords=[3]{
      forall,
    },
    keywordsprefix=\#,
    keywords=[4]{
      true,
      false,
      null
    },
    keywords=[5]{
      doc,
      default,
    },
    sensitive=true, 
    morecomment=[l]{//}, 
    morecomment=**[is][\color{gray}]{\$}{\$},
    morestring=[b]", 
    morestring=[s]{m\#"}{"\#m},
    moredelim=[is][\color{gray}]{\$}{\$},
    literate=
        *{->}{$\rightarrow$}1
        {...}{\ldots}1
        {@&}{$\cap$}1
        {@|}{$\cup$}1
        {=>}{$\Rightarrow$}1
        {>=}{$\geq$}1
        {<=}{$\leq$}1
} %

\lstdefinelanguage{Racket} {
  morekeywords=[1]{define, define-syntax, define-macro, lambda, define-stream, stream-lambda},
  morekeywords=[2]{begin, call-with-current-continuation, call/cc,
    call-with-input-file, call-with-output-file, case, cond,
    do, else, for-each, if,
    let*, let, let-syntax, letrec, letrec-syntax,
    define-context, define-controller, Integer, Boolean, get, when-required, when-provided,
    maybe_publish, require, submod, or/c, and/c, ->, \#\%module-begin,
    always_publish, with-syntax, define-struct/contract, syntax-case,
    define/contract,
    let-values, let*-values,
    module, provide,
    and, or, not, delay, force,
    \#`, \#',
    \#lang, implement, begin-for-syntax, rename-out,
    quasiquote, quote, unquote, unquote-splicing,
    map, fold, syntax, syntax-rules, eval, environment, query },
  morekeywords=[3]{import, export},
  alsoletter={',`,-,/,>,<,\#,\%},
  morecomment=[l]{;},
  moredelim=**[is][\color{light-gray}]{<<@<<}{>>@>>},
  moredelim=**[is][\itshape\color{OliveGreen}]{<<;<<}{>>;>>},
  morecomment=[s]{\#|}{|\#},
  sensitive=true,
}

\lstdefinelanguage{JavaScript}{
  keywords={typeof, new, true, false, catch, function, return, null, catch, switch, var, if, in, while, do, else, case, break},
  ndkeywords={class, export, boolean, throw, implements, import, this},
  ndkeywordstyle=\color{darkgray}\bfseries,
  identifierstyle=\color{black},
  sensitive=false,
  comment=[l]{//},
  morecomment=[s]{/*}{*/},
  commentstyle=\color{purple}\ttfamily,
  morestring=[b]',
  morestring=[b]"
}

\lstdefinelanguage{Python}{
  identifierstyle=\color{black},
  sensitive=false,
  morestring=[b]',
  morestring=[b]"
}

\newcommand{\nickel}[1]{\lstinline[language=nickel]{#1}}
\newcommand{\racket}[1]{\lstinline[language=racket]{#1}}

\newcommand{\typescript}[1]{\lstinline[language=JavaScript]{#1}}
\newcommand{\python}[1]{\lstinline[language=Python]{#1}}

\begin{abstract}
Union and intersection types are a staple of gradually typed languages such
as TypeScript. While it's long been recognized that union and
intersection types are difficult to verify statically, it may appear
at first that the dynamic part of gradual typing is actually pretty
simple.

It turns out however, that in presence of higher-order contracts union and
intersection are deceptively difficult. The literature on higher-order contracts
with union and intersection, while keenly aware of the fact, doesn't really
explain why. We point and illustrate the problems and trade-offs inherent to
union and intersection contracts, via example and a survey of the literature.
\end{abstract}

\title{Union and Intersection Contracts Are Hard, Actually}
\author{Teodoro Freund}
\affiliation{
  \institution{Universidad de Buenos Aires}
  \city{Buenos Aires}
  \country{Argentina}
}
\email{tfreund@dc.uba.ar}

\author{Yann Hamdaoui}
\affiliation{
  \institution{Tweag}
  \city{Paris}
  \country{France}
}
\email{yann.hamdaoui@tweag.io}
\author{Arnaud Spiwack}
\affiliation{
  \institution{Tweag}
  \city{Paris}
  \country{France}
}
\email{arnaud.spiwack@tweag.io}

\setcopyright{acmlicensed}
\acmPrice{15.00}
\acmDOI{10.1145/3486602.3486767}
\acmYear{2021}
\copyrightyear{2021}
\acmSubmissionID{splashws21dlsmain-p31-p}
\acmISBN{978-1-4503-9105-4/21/10}
\acmConference[DLS '21]{Proceedings of the 17th ACM SIGPLAN International Symposium on Dynamic Languages}{October 19, 2021}{Chicago, IL, USA}
\acmBooktitle{Proceedings of the 17th ACM SIGPLAN International Symposium on Dynamic Languages (DLS '21), October 19, 2021, Chicago, IL, USA}

\begin{document}

\ccsdesc[300]{General and reference~Surveys and overviews}
\ccsdesc[500]{Software and its engineering~Language features}
\ccsdesc[300]{Software and its engineering~Software verification and validation}

\keywords{contracts, higher-order contracts, union, intersection}

\maketitle

\section{Introduction}
\label{sec:intro}

Union types, meaning a type \nickel{A @| B} containing values which
belong either to a type \nickel{A} or \nickel{B}, are a popular tool
when adding static types to a dynamic language. In particular, both
TypeScript~\cite{TypeScriptUnions} and MyPy~\cite{MyPyOptional}, use
union types to model the frequent practice to use the value
\typescript{null} (\python{None} in Python) to represent an absent
optional value. This is why the gradual typing literature, concerned
with formalising the interplay between static and dynamic type
systems, has been quite interested in union
types~\cite{RootCauseOfBlame,gradualCastagna,ORTIN2011278,ToroTanterGradualUnion,KeilThiemannUnionIntersection}.

On the other hand, unions are not a common feature of static type
systems, mostly because they are quite difficult to verify
statically. So unions are really only worth it in gradually typed
language where they formalise existing dynamically typed patterns.

Surely, for dynamic tests, unions ought to be really easy: they
are simply the Boolean disjunction of two dynamic tests, that fail
whenever one of those tests fail.
Unfortunately, as we document in this article, as soon as you extend
dynamic checks to \emph{contracts}~\cite{FindlerFelleisenHOContracts},
unions become actually pretty difficult, and threaten desirable
properties of your language.

\subsection{Configuration Languages}

To motivate contracts and the problem caused by unions, let's make a
detour through configuration languages.
A configuration language is a language concerned with describing the
configuration of an application. In traditional configuration
languages, such as YAML, TOML, or JSON, the configuration is fully,
and explicitly, spelt out.

However, with the advent of DevOps, configurations have been extended
to describe the entire state of a computer, or even a fleet of
computers. For instance, with Kubernetes you need to configure a large
fleet of (possibly replicated) docker containers. To describe this
sort of configurations, you really want to be able to re-use and
abstract parts of the configuration, like traditional programming
languages do. To meet this need, languages such as Cue~\cite{cueLang},
Dhall~\cite{dhallLang}, Jsonnet~\cite{jsonnetLang}, or Nickel~\cite{NickelRepo}, where
configurations are generated rather than spelt out, were created.

Another example is continuous integration systems: it's fairly typical
to need a matrix of jobs, wherein the same tests are run on different
infrastructures, or with different versions of a compiler. Traditional
configuration would have you copy the same steps for each
infrastructure. This is tedious, hard to maintain, and error
prone. It's much better, instead, to write the steps once, and
instantiate them for each infrastructure. Continuous integration
systems typically do this using a templating system layered on top of
YAML. Each of the configuration-generating languages above allow such
job-matrix definition natively.

\subsection{Nickel}

In this article, we will use the Nickel language~\cite{NickelRepo} as
illustration and motivation. At its core, Nickel is the JSON data
model, augmented with abstraction mechanisms, and it comprises:

\begin{itemize}
  \item dictionaries, written as
    \footnote{Note that, unlike JSON, Nickel assigns values in
    dictionaries using \nickel{=}, to keep \nickel{:} for type
annotations.}:
\begin{lstlisting}[frame=none,numbers=none,language=nickel]
{field1 = value1, ..., fieldn = valuen}
\end{lstlisting}
  \item arrays:
\begin{lstlisting}[frame=none,numbers=none,language=Nickel]
[x1, x2, ..., xn]
\end{lstlisting}
  \item functions:
\begin{lstlisting}[frame=none,numbers=none,language=Nickel]
fun arg1 ... argn => body
\end{lstlisting}
  \item and let-definitions:
\begin{lstlisting}[frame=none,numbers=none,language=Nickel]
let id = value in exp
\end{lstlisting}
\end{itemize}

A Nickel configuration is then evaluated to an explicit configuration,
\emph{e.g.}  in JSON, which can then be consumed by an
application. Therefore a design constraint of Nickel is any Nickel
data must have a straightforward interpretation in JSON.

\subsection{Contracts}
\label{sub-sec:contracts}

A useful feature of a configuration language is to provide facilities
for schema validation. That is, help answer questions like: does our
configuration have all the required fields? does the \nickel{url} field indeed
contains a URL?

These are inherently dynamic questions, as they are all questions
about the evaluated configuration. To this effect, Nickel lets us
annotate any expression with a dynamic schema check: \nickel{exp |
  C}. There is also syntactic sugar to annotate definitions:
\nickel{let id | C = value in exp} stands for \nickel{let id = (value
  | C) in exp}.

Let us pause for a moment and consider the following: it is Nickel's
ambition to be able to manipulate configurations like Nixpkgs. With
over 50\,000 packages, it is one of the largest repository of software
packages in existence~\cite{repology}. Concretely, Nixpkgs is a
dictionary mapping packages to build recipes. That is, a massive,
over-50\,000-key-value-pair wide dictionary. It is absolutely out of the
question to evaluate the entirety of this dictionary every time one
needs to install 10 new packages: this would result in a painfully
slow experience.

To be able to support such large dictionaries, Nickel's dictionaries
are \emph{lazy}, that is, the values are only evaluated when
explicitly required. For instance, when evaluating the expression
\hbox{\nickel{nixpkgs.hello},} only the \nickel{hello} package gets
evaluated, even if \nickel{nixpkgs} contained a
\nickel{world} package as well.

But let's consider now writing something like \nickel{nixpkgs |
  packages}, to guarantee that all the packages conform to the desired
schema. If this were a simple Boolean test, it would have to evaluate
all 50\,000 package to check their validity, hence breaking the
laziness of dictionaries. Do we have to choose between laziness and
schema validation? Fortunately, we don't! Enter
\emph{contracts}~\cite{FindlerFelleisenHOContracts}: dynamic checks
which can be partially delayed, yet errors can be reported
accurately. Contracts can respect laziness of dictionaries, and they
can be used to add schema validation to functions as well (in fact
functions were the original motivation for contracts).

There is no Boolean function which can check that a value has type
\nickel{Str -> Str}. Instead, a contract for \nickel{Str -> Str} \\
checks for each call of the function whether
\begin{enumerate}
\item the argument has type \nickel{Str}, otherwise the caller of the
  function is faulty
\item if so, that the returned value has type \nickel{Str}, otherwise
  the implementation of the function is faulty
\end{enumerate}

Like in the case of lazy dictionaries, the checks are delayed.
Contracts keep track of whether the caller or the implementation is
at fault for a violation, hence it can report precise error
messages. Contracts are said to \emph{blame} either the caller or the
implementation. Compare Figure~\ref{fig:contract-reporting-wo} and
Figure~\ref{fig:contract-reporting-w}: in
Figure~\ref{fig:contract-reporting-wo} an error is reported inside the
\nickel{catHosts} function, but \nickel{catHosts} is, in fact,
correct, as is made clear by Figure~\ref{fig:contract-reporting-w},
where \nickel{catHosts} is decorated with the \nickel{Str -> Str}
contract, and correctly reports that the caller failed to call
\nickel{catHosts} with a string argument.

\begin{figure*}
  \centering
  \begin{subfigure}[b]{0.48\linewidth}
    \begin{lstlisting}[language=Nickel]
let catHosts = fun last =>
  let hosts = ["foo.com", "bar.org"] in
  lists.fold (fun val acc =>
    val ++ "," ++ acc) hosts last in

let makeHost = fun server ext =>
  server ++ "." ++ ext in

catHosts (makeHost "google")
\end{lstlisting}

    \begin{lstlisting}[frame=none,numbers=none, basicstyle=\footnotesize\ttfamily]
error: Type error
3 | [(*@{\ldots}@*)] "," ++ acc) hosts last in
  |            ^^^
  | This expression has type Fun,
  | but Str was expected
4 |
5 | let makeHost = fun server ext => [(*@{\ldots}@*)] in
  |                --------------------
  |                evaluated to this
  = ++, 2nd argument
\end{lstlisting}
    \caption{Error reporting without contract}
    \label{fig:contract-reporting-wo}
  \end{subfigure}
  \hfill
  \begin{subfigure}[b]{0.48\linewidth}
    \begin{lstlisting}[language=nickel]
$let$ catHosts | Str -> Str $= fun last =>
  let hosts = ["foo.com", "bar.org"] in
  lists.fold (fun val acc =>
    val ++ "," ++ acc) hosts last in

let makeHost = fun server ext =>
  server ++ "." ++ ext in

catHosts (makeHost "google")$
\end{lstlisting}
    \begin{lstlisting}[frame=none,numbers=none, basicstyle=\footnotesize\ttfamily]
error: Blame error: contract broken by the caller.
  |  Str -> Str
  |  --- expected type of the argument [(*@{\ldots}@*)]
[(*@{\ldots}@*)]
1 | let catHosts | Str -> Str = fun last =>
  |                ^^^^^^^^^^ bound here
[(*@{\ldots}@*)]
6 | catHosts (makeHost "google")
  | --------------------------- (2) calling <func>
    \end{lstlisting}
    \caption{Error reporting with contract}
    \label{fig:contract-reporting-w}
  \end{subfigure}
  \caption{Contracts improve error messages}
\end{figure*}

As we shall show, the delayed check of contract, while essential to
ensuring that schema validation doesn't affect performance (or, indeed,
is possible at all on functions), make union contracts (and their less
appreciated sibling, intersection contracts) quite problematic.
While usual contracts require only one witness to show the invalidation
of a contract, the introduction of unions makes
the number of witnesses not bounded.

\subsection{Contributions}
Our contributions are as follows
\begin{itemize}
\item We describe the fundamental difficulties caused by pre-sence of
  union and intersection contracts in a language, which are kept
  implicit in the literature (Section~\ref{sec:issues-sem})
\item We survey the various trade-offs which appear in implemented
  languages and in the academic literature to work around these difficulties
  (Section~\ref{sec:issues-literature})
\end{itemize}

\section{A Typology of Language Features}
\label{sec:feat-lang}

Union contracts are not only difficult to implement, their
unrestricted presence is incompatible with potentially desirable
properties of the language. In this section we present some of these
properties; we will show how these properties interact with union
contracts in Sections~\ref{sec:issues-sem}
and~\ref{sec:issues-literature}.

\subsection{User-Defined Contracts}
\label{sec:flat-contracts}

A strength of dynamic checking is that we can easily check properties which
are impractical to check statically. For instance that a string
represents a well-formed URL, or a number is a valid port.

This same property is desirable of contracts as well, otherwise we
lose an important benefit of dynamic checking. Preferably, we want to
be able to extend the universe of contracts with user-defined
predicates.

For instance, Figure~\ref{fig:port-contract} shows the definition of a
contract for valid ports in Nickel syntax.
User-defined contracts can be combined with other contracts normally:
\nickel{Int -> Port} is a contract verified by functions
which, given an integer returns a valid port.

\begin{figure}[h]
  \begin{center}
\begin{lstlisting}[language=Nickel]
let Port = contracts.fromPred (fun p =>
  num.isInt p && 0 <= p && p <= 65535) in
80 | Port
\end{lstlisting}
\end{center}
\caption{A contract for valid ports}
\label{fig:port-contract}
\end{figure}

This type of contracts are present in many different languages,
for instance, the Eiffel programming language~\cite{meyer1987eiffel}, the precursor
of the Design by Contract philosophy, makes it possible to assert
these kinds of expression as pre- and post-conditions on
functions and as invariants on classes~\cite{EiffelDesignByContract}.

The Racket programming language also has a system to work with
contracts, powerful enough to define user-defined contracts, and
to compose them with other kinds of dynamic checks,
like higher order contracts or a lightweight take on union
and intersection contracts~\cite{RacketContracts}.

\subsection{Referential Transparency}
\label{sec:optimizations}

The performance of modern programs heavily relies on the optimizations performed
by the compiler or the interpreter. Even more so for functional languages, whose
execution model is often far removed from the hardware, causing naive execution
to exhibit unacceptable slowdowns.

One such important optimization is inlining (Figure
\ref{fig:optimizations-inlining-ex}). Functional programs tend to make heavy use
of functions, and a function call is not a free operation: it usually involves a
number of low-level operations such as saving and storing registers, pushing a new stack
frame and jumping to and back from the function's body. Inlining eliminates a
function call by directly substituting the function for its definition at
compile time (or before execution, for interpreted language). This is especially
efficient for small functions that are called repeatedly.

\begin{figure}[h]
  \begin{center}
\begin{lstlisting}[language=Nickel,title={Source program}]
let elem = fun elt =>
  lists.any (fun x => x == elt) in

let subList = fun l1 l2 =>
  elem (lists.head l1) l2
  && subList (list.tail l1) l2
\end{lstlisting}
\begin{lstlisting}[language=Nickel,title={Optimized program}]
let subList = fun l1 l2 =>
  lists.any (fun x => x == (lists.head l1))
            l2
  && subList (list.tail l1) l2
\end{lstlisting}
  \end{center}
\caption{Inlining}
\label{fig:optimizations-inlining-ex}
\end{figure}

While inlining expands an expression by substituting a definition for its value,
an opposite transformation can be beneficial when a composite expression occurs
at several places. In this case, the same expression is wastefully recomputed at
each occurrence. Common subexpression elimination (CSE) consists in storing the
expression in a variable that is then used in place of the original occurrences
(Figure \ref{fig:optimizations-cse-ex}), thus evaluating the expression once and
for all.

\begin{figure}[h]
  \begin{center}
\begin{lstlisting}[language=Nickel,title={Source program}]
let elemAtOrLast = fun index list =>
  if index > lists.length list - 1 then
    lists.elemAt (lists.length list - 1)
                 list
  else
    lists.elemAt index list
\end{lstlisting}
\begin{lstlisting}[language=Nickel,title={Optimized program}]
let elemAtOrLast = fun index list =>
  let l = lists.length list - 1 in
  if index > l then
    lists.elemAt l list
  else
    lists.elemAt index list
\end{lstlisting}
  \end{center}
\caption{Common subexpression elimination}
\label{fig:optimizations-cse-ex}
\end{figure}

Beyond CSE, other optimizations such as loop-invariant code motion or
let-floating~\cite{letFloating} apply the same principle of extracting out an
invariant expression to avoid recomputing it (respectively across loop
iterations and function calls).

\begin{figure}
  \begin{center}
\begin{lstlisting}[language=Nickel,title={Source program}]
let f = fun x => g y (x + 1)
\end{lstlisting}
\begin{lstlisting}[language=Nickel,title={Optimized program}]
let g' = g y in
let f' = fun x => g' (x + 1)
\end{lstlisting}
  \end{center}
\caption{Let-floating}
\label{fig:optimizations-let-floating-ex}
\end{figure}

For example, take the code of Figure~\ref{fig:optimizations-let-floating-ex}.
The partial application \lstinline+g y+ is recomputed each time \lstinline+f+ is
called. This may be costly, in particular in the presence of contracts: if the
first argument of \lstinline+g+ must be a list with elements of a specific kind
and if that precondition is enforced by a function contract (say \nickel{g |
List Odd -> Even -> List Odd}), the additional cost is linear in the size of
\lstinline+y+. A sensible thing to do is to factor \lstinline+g y+ out of
\lstinline+f+ as in Figure~\ref{fig:optimizations-let-floating-ex}, which is
something a let-floating transformation could indeed do (given \lstinline+g+ is
pure, as detailed below).

The soundness of these optimizations is tied to the validity of specific program
equivalences. Inlining requires that one can replace the application of a
function by its body, which is basically $\beta$-reduction: as long as the
arguments are evaluated following the language's strategy, this is usually a
valid transformation. However, Section~\ref{sec:issues-sem} exposes that the
question of inlining a function with a contract attached is more subtle.

A CSE-like transformation on a term $M$ requires on the other hand an
equivalence of the form:

\begin{equation}\label{eq:cbn-expansion}
M[N/x] \simeq let~x~=~N~in~M
\end{equation}

$M[N/x]$ stands for the substitution of $x$ for the term $N$ in the term $M$.
This equation clearly fails in presence of side-effects, as demonstrated in
Figure~\ref{fig:optim-invalid-cse}. In that example, \nickel{(f 1, f 1)} prints
\nickel{"hi"} two times while \nickel{let y = f 1 in (y,y)} only prints it once.

\begin{figure}[h]
\begin{lstlisting}[language=Nickel,title={Effectful function}]
let f x = print "hi";(x+1)
\end{lstlisting}
\begin{lstlisting}[language=Nickel,title={Invalid expansion}]
(f 1,f 1) (*@ $\not \simeq$ @*) let y = f 1 in (y,y)
\end{lstlisting}
\caption{Counter-example to (\ref{eq:cbn-expansion}) in presence of
side-effects}
\label{fig:optim-invalid-cse}
\end{figure}

However, (\ref{eq:cbn-expansion}) does hold for \emph{pure} terms, that are
terms without side-effects. In a pure language, all these transformations are
valid.  In impure languages, the situation varies: in some case a large subset
of pure terms can be identified (in languages with effects tracking such as
PureScript) to be safely transformed.  Otherwise, the compiler must stay
conservative and only apply CSE to expressions it can prove are without
side-effects (arithmetic expressions, for example).

Strikingly, we will see in Section~\ref{sec:issues-sem} that the introduction of
union and intersection contracts \emph{breaks referential transparency} and make
(\ref{eq:cbn-expansion}) \emph{unsound}\footnote{(\ref{eq:cbn-expansion}) holds
  in Nickel despite non-termination and contract-checking errors
  because of lazy evaluation. In strict languages (\ref{eq:cbn-expansion}) needs to be restricted, nevertheless
  union and intersection contracts make it worse.}, preventing the kind of optimization of
Figure~\ref{fig:optimizations-let-floating-ex} to fire in general.

\section{Union \& Intersection}
\label{sec:union-inter}

Let us now consider union and intersection contracts, before we explain in
Section~\ref{sec:issues-sem} how they can compromise the properties that we
described in Section~\ref{sec:feat-lang}.

\subsection{Unions}

A union type \nickel{A @| B} is a type of values which are either of
type \nickel{A} or of type \nickel{B}: literally the union of
\nickel{A} and \nickel{B}. Union types are popular in gradual typed
systems such as TypeScript~\cite{TypeScriptUnions} and
MyPy~\cite{MyPyOptional}.

\paragraph{In Gradually Typed Systems}

The problem that these practical gradual type systems are trying to solve is to
capture, in static types, as many programming patterns as possible
from the underlying dynamically typed language (Java\-Script for
TypeScript and Python for MyPy). One such pattern is
heterogeneous collections. For instance, in TypeScript, an array
which can contain both strings and numbers would have type
\typescript{Array<string|number>}.

A probably even more common pattern is a variable which can contain
either a value of type \typescript{A} (say, a number) or the \typescript{null}
value. So much so, in fact, that MyPy defines a type alias
\python{Optional[A]} for \python{Union[A,None]} (\python{None} is how
Python renders the null value).

Yet another application of union types is, rather than capturing a
pattern from JavaScript or Python, to capture a pattern from
traditional statically typed language: sum types. In statically typed
languages, values of sum types are usually thought of as being built
out of constructors. But neither JavaScript nor Python have such
constructors. So instead, sums are construed as ``tagged unions'' (or
discriminated unions), that is, quite literally, the union of two
types which contain a discriminating tag. See
Figure~\ref{fig:tagged-union} for an example from the TypeScript
documentation: there the \typescript{kind} field is the tag, and its
type in both alternatives is a singleton type which contains only the
specified string.

\begin{figure}
  \centering
  \begin{lstlisting}[language=JavaScript]
    interface Circle {
      kind: "circle";
      radius: number; }

    interface Square {
      kind: "square";
      sideLength: number; }

    type Shape = Circle | Square;
\end{lstlisting}

  \caption{A sum type as a tagged union}
  \label{fig:tagged-union}
\end{figure}

\paragraph{Union contracts}

In the academic gradual type literature, it is common to use contracts
as a glue between static and dynamic types. Therefore, the question of
bringing union to contracts is natural, and have indeed been studied
(\emph{e.g.}~\cite{KeilThiemannUnionIntersection,RootCauseOfBlame}).

Like for static types, a value which satisfies the contract \nickel{A @| B}
is a value which satisfies either contract \nickel{A} or
contract \nickel{B} (though in Section~\ref{sec:issues-literature} we
will see that it may be desirable to weaken this definition).

Nickel is a language built from scratch with contracts, so it may be
less clear why unions are useful. However, Nickel's ambition is to
have its data model canonically interpretable in common serialization
formats, in particular JSON. It means that it is very convenient to
represent optional value by the \nickel{null} value like in
JavaScript. It also means that Nickel doesn't have built-in
constructors: constructors don't have a canonical representation in
JSON. So it would be quite natural to represent optional contracts and
sum contracts as unions.

\subsection{Intersections}

An intersection type \nickel{A @& B} is satisfied by values which
satisfy both contract \nickel{A} and contract \nickel{B}.

Intersection contracts (and types) are probably less prevalent than union in
practical type systems. However, a function from a union is equivalent to an
intersection. That is \nickel{(A @| B) -> C} $\simeq$ \nickel{(A -> C) @& (B ->
C)}. So in a system with functions and unions, intersections are already morally
present (and, for that matter, in a system with functions and intersections,
unions are morally present). Some of our examples in
Sections~\ref{sec:issues-sem} and~\ref{sec:issues-literature} are better
expressed in terms of intersections, so it's best to include them.

Figure~\ref{fig:addElem} gives a concrete example of this phenomenon. The
function \nickel{appendDate} appends an element to \nickel{list}, whose type is
only known to be the union of two lists, each using a different representation.
Because the return type is the same as the input type, \nickel{appendDate} must
preserve this representation (\nickel{Date} and \nickel{DateWeek}
cannot be mixed in a same list).  Both alternatives support falling back to
a simple string for unparsed dates.  Faced with these two possibilities,
\nickel{appendDate} can only append a value which fits both types: this is
precisely the intersection \nickel{(Date @| Str) @& (DateWeek @| Str)}, that
is, \nickel{Str}.

\begin{figure}[h]
\begin{lstlisting}[language=Nickel]
let Date = {day | Num, month | Num, year | Num} in
let DateWeek = {dayOfWeek | Num, week | Num, year | Num} in
let appendDate | (List (Date @| Str)
                 @| List (DateWeek @| Str))
                 -> (List (Date @| Str)
                 @| List (DateWeek @| Str)) =
  fun list => lists.cons "01/01/2021" list
\end{lstlisting}
\caption{Adding an element to a union of two arrays}
\label{fig:addElem}
\end{figure}

Yet, intersection are useful in their own right: they can be used to
combine dictionaries in the style of object-oriented multiple inheritance. For instance, in
Figure \ref{fig:intersection-record} two types are defined
\nickel{Animal} and \nickel{Pet}, and a variable that is compatible
with both types is declared, with type \nickel{Animal @& Pet}. This
particular application is supported, for instance, by TypeScript.

\begin{figure}[h]
\begin{lstlisting}[language=Nickel]
let Animal =
  { species | Str, breed | Str, name | Str } in
let Pet = { owner | Str, name | Str } in
let myDog | Animal @& Pet =
  { species = "Canis Lupus",
    breed   = "Australian Cattle Dog",
    owner   = "Anonymous Author",
    name    = "Juno" }
\end{lstlisting}
\caption{An animal that is also a pet}
\label{fig:intersection-record}
\end{figure}

Another application of intersections shows up when intersecting
functions: it can be used to encode overloading.
For instance, take a look at Figure~\ref{fig:intersection-overloading-ex}, where the function
\nickel{duplicate} works both as a function to duplicate
arrays, as well as a function to duplicate strings.
This is particularly useful when using unions, since it's a
good way to express that a function can deal with different
shapes of data.
For instance, in the same figure, \nickel{duplicate} is (correctly) called
on a value of type \nickel{(List Str) @| Str}.

\begin{figure}[h]
\begin{lstlisting}[language=Nickel]
let duplicate
    | (List Str -> List Str)
      @& (Str -> Str) =
  fun x => x ++ x in
let text | (List Str) @| Str = ... in
duplicate text
\end{lstlisting}
\caption{Duplicating an array of Strings or a String}
\label{fig:intersection-overloading-ex}
\end{figure}

\section{Incompatibilities}
\label{sec:issues-sem}

However appealing union and intersection contracts may be, they happen to be
either hard to combine or even fundamentally incompatible with the desirable
language features from Section~\ref{sec:feat-lang}. At least in their full-blown
form: in Section~\ref{sec:issues-literature} we will discuss pragmatic
restrictions of union and intersection contracts to recover some or all of the
features.

\subsection{Union Contracts as a Side-Effect}

In Nickel, the failure of a function contract can always be traced back to a
single call. For example, take the function \nickel{f} with a simple contract
attached of Figure~\ref{fig:pos-to-pos}. The whole program fails with a contract
error blaming \nickel{f} because the return value of the second call \nickel{f
5} violates the \nickel{Positive} contract. The first call to \nickel{f} does
not matter, and \nickel{f 5} is a single and independent witness of the contract
violation. The user is pointed to this one location in practice.

This single witness property can be justified as follows. Apart from the error
reporting part (although this is the crucial bit in practice!), the current
contract system of Nickel can be implemented purely as a library, requiring only
a \nickel{fail} primitive to abort the execution. In practice, applying a
function contract to \nickel{f} replaces it with an \nickel{f'} that performs the
additional checks. Thus, since the core language is pure (albeit
partial, if only because \nickel{fail}), the failure of \nickel{f' 5}
must be independent of its environment and of any previous call to \nickel{f'}.

%

\begin{figure}[h]
\begin{lstlisting}[language=Nickel]
let f | Positive -> Positive
      = fun x => x - 7 in
(f 10) + (f 5)
\end{lstlisting}
\caption{Simple contract violation}
\label{fig:pos-to-pos}
\end{figure}

%

Union contracts are different. Consider the program presented in
Figure~\ref{fig:wrong-union-function}. The same \nickel{f} is now given a union
contract. \nickel{f} is violating this contract once again, as it neither maps
all positive numbers to positive numbers nor to nonpositive numbers.

\begin{figure}[h]
\begin{lstlisting}[language=Nickel]
let f | (Positive -> Positive)
        @| (Positive -> NonPositive)
      = fun x => x - 7 in
(f 10) + (f 5)
\end{lstlisting}
\caption{Union contract violation}
\label{fig:wrong-union-function}
\end{figure}

This program must fail, because \nickel{f 10} is a witness of \nickel{f} failing
the contract \nickel{Positive -> NonPositive}, and \nickel{f 5} is a witness of
\nickel{f} failing \nickel{Positive -> Positive}.  But, as opposed to the
example from Figure~\ref{fig:pos-to-pos}, removing only one of the calls makes the
program succeed! Indeed, each call only unveils the violation of one component
of the union. In this example, a single call to \nickel{f} that would be the
witness of the violation of the whole contract doesn't even exist: a minimum of
two are always needed.

This behavior shows that union contracts introduce side-effects. The result
of \nickel{f 5} now depends on the previous execution and more specifically on
any prior call to \nickel{f}. This behavior of union contracts breaks
referential transparency, as well as the property~\ref{eq:cbn-expansion}
introduced in Section~\ref{sec:optimizations}, that is required to perform
CSE-like optimizations.

Figure~\ref{fig:optimized-programs} illustrates this point further. It contains
an original program and an optimized version where the common subexpression
\nickel{f 1} has been eliminated. While equivalent in a pure language with only
non-termination or plain higher-order contracts, these two programs behave
differently because of unions:

\begin{itemize}
    \item The original version returns \nickel{(1, "False")} without failing.
    \item The optimized version fails with a contract violation.
\end{itemize}

In the original version, each partial application \nickel{f 1} gives rise to a
fresh instance of the contract \nickel{Bool -> Num @| Bool -> Str}. These
instances are independent, and can pick a different component of the union to
satisfy. Although \nickel{f} doesn't actually respect the contract, these calls
are not enough to prove so. In the optimized version, \nickel{g} is endowed with
a single contract, that must pick one of the two components of the union. There,
the two calls refer to the same union contract, and shows that \nickel{f} does
violate its initial contract.

\begin{figure}[h]
\begin{lstlisting}[language=Nickel, title=Original]
let f | Num -> (Bool -> Num @| Bool -> Str)
      = fun x y => if y then x else "False"
in (f 1 true, f 1 false)
\end{lstlisting}
\begin{lstlisting}[language=Nickel, title=Optimized]
let f | Num -> (Bool -> Num @| Bool -> Str)
      = fun x y => if y then x else "False"
let g = f 1 in
(g true, g false)
\end{lstlisting}
\caption{Equivalent programs with CSE applied}
\label{fig:optimized-programs}
\end{figure}

To sum up, the addition of union contracts introduce side-effects in a pure
language. Side-effects have well-known pitfalls:
\begin{itemize}
    \item For the programmer, they are hard to reason about. They prevent local
        reasoning. In our previous examples, removing or adding a function call
        somewhere can toggle a failure in a call at a totally different
        location.
    \item For the interpreter (or compiler), side-effects inhibit many optimizations and
        program transformations.
\end{itemize}

%

\subsection{Intersection with User-Defined Contracts}
\label{sec:flat-and-inter}

A natural --- but naive --- implementation of intersection contracts could be the
following: to apply a contract \nickel{A @& B}, apply both contracts \nickel{A}
and \nickel{B} sequentially, resulting in the naive decomposition rule of
Figure~\ref{fig:naive-impl}.

\begin{figure}[h]
\begin{lstlisting}[language=Nickel,frame=none,numbers=none,title={Naive
decomposition}]
M | A @& B (*@ $\simeq$ @*) (M | A) | B
\end{lstlisting}
\begin{lstlisting}[language=Nickel,frame=none,numbers=none,title={Exchange law}]
(A -> B) @& (C -> D) (*@ $\simeq$ @*) (A @& C) -> (B @& D)
\end{lstlisting}
\begin{lstlisting}[language=Nickel,title={Overloaded identity}]
let g | Num -> Num @& Str -> Str
      = fun x => x in
g 1
\end{lstlisting}
\caption{Naive implementation of intersection}
\label{fig:naive-impl}
\end{figure}

This intuition works for simple contracts: checking that \nickel{x | Natural @&
Odd} amounts to check that \nickel{x | Natural} and \nickel{x | Odd}.
Unfortunately, this doesn't scale to higher-order contracts.  The overloaded
identity example of Figure~\ref{fig:naive-impl} illustrates the use of an
intersection to model a simple overloading of the identity function. If we were
to apply the naive decomposition, the argument \nickel{1} would fail the
\nickel{Str -> Str} contract and abort the execution. Perhaps the exchange rule
given in Figure~\ref{fig:naive-impl}, which is a direct consequence of the naive
decomposition, illustrates the issue better. It is clear that this exchange law
isn't the right semantics for overloading. With this law, the contract for
overloaded identity of Figure~\ref{fig:naive-impl} would always fail because
no argument can satisfy \nickel{Num @& Str}.


In a higher-order intersection contract, blame is raised when:
\begin{description}
    \item[Faulty caller] The argument fails \emph{both} components.
    \item[Faulty implementation] The function fails at least \emph{one}
        component that the argument previously satisfied.
\end{description}


%

To fix the naive implementation, the interpreter can share state between the
sub-contracts, in order to decide if blame must be raised or not when a
sub-contract fails:

\begin{lstlisting}[language=Nickel,frame=none,numbers=none]
x | A @& B (*@ $\simeq$ @*) (x | A[l]) | B[l]
\end{lstlisting}

Shared state is represented by the label \nickel{l}. Such a shared state is in essence the
approach proposed by Williams, Morris, and Wadler in~\cite{RootCauseOfBlame}.



\begin{figure}[h]
\begin{lstlisting}[language=nickel]
let C = contracts.fromPred (fun f =>
  f 0 == 0) in
let g | (Str -> Str) @& C
      = fun x => x
in g 0
\end{lstlisting}
\caption{Intersection and user defined contracts}
\label{fig:inter-flat-contracts}
\end{figure}

However, this shared-state approach has a major drawback: it isn't straightforwardly
compatible with user-defined contracts (introduced in Section
\ref{sec:flat-contracts}). The issue is similar to our initial issue with higher-order
contracts and the naive decomposition: user-defined contracts may apply
functions and thus make a sub-contract of the intersection fail, but this
failure shouldn't always result in raising blame. An example is given on Figure
\ref{fig:inter-flat-contracts}. Decomposing using the shared state approach, we
end up with:

\begin{lstlisting}[language=Nickel,frame=none,numbers=none,title={Stateful
decomposition}]
((fun x => x) | (Str -> Str)[l]) | C[l]
\end{lstlisting}
where \nickel{l} represents the shared state.  At this point,
applying the \nickel{C} contract results in evaluating:

\begin{lstlisting}[language=Nickel,frame=none,numbers=none]
((fun x => x) | (Str -> Str)[l]) 0 == 0.
\end{lstlisting}

Applying a function wrapped in a \nickel{Str -> Str} contract to \nickel{0}
fails negatively. This is not the expected behavior, since the identity function
does respect semantically both contracts. As opposed to built-in higher-order
contracts, user-defined contracts are black-box from the interpreter's point of
view, and it is thus not obvious how to extend the shared state approach to
handle user-defined contracts.

Once again, intersection contracts introduce side-effects in the picture. What's
more, these side-effects interact with user-defined contracts in a non-trivial
way, while they are an important feature for validation.

\section{Pragmatic Trade-Offs}
\label{sec:issues-literature}

Despite the difficulties of Section~\ref{sec:issues-sem}, union and
intersection contracts are still sought after. In this section we turn
to existing systems with union and intersection contracts in the
literature and in implementations.

These systems all make trade-offs, sacrificing some features of union
and intersection contracts to preserve language features. We survey
and discuss those trade-offs and their implications.

\subsection{A Coinductive Semantics}
\label{sec:coinductive-sem}

In order to give a precise definition to what values ought to satisfy
union and intersection contracts,
\citeauthor{KeilThiemannUnionIntersection}~\cite{KeilThiemannUnionIntersection}
give a coinductively defined semantics inspired by union and
intersection type systems.
The key innovation of their work is recognizing that giving a
semantics to higher-order contracts requires defining not only what
values satisfy a contract, but also what \emph{contexts} satisfy the
contract. This models the situation where context may violate a
contract by calling a function with an inappropriate argument.

Concretely, given a contract $C$, \citeauthor{KeilThiemannUnionIntersection}
introduce the two sets $\llbracket C \rrbracket^+$ and $\llbracket C
\rrbracket^-$ of values and contexts, respectively, satisfying the contract.
They are defined by mutual induction and coinduction.

This semantics has limited support for overloading. Consider the example in
Figure~\ref{fig:intersection-distribution}: it could evaluate to the pair
\nickel{(1, 1)}. But \citeauthor{KeilThiemannUnionIntersection}'s coinductive
semantics rejects it as a contract violation.  This can be phrased pithily as
the fact that the coinductive semantics doesn't validate the property
\nickel{A -> B @& A -> C}$\simeq$\nickel{A -> (B @& C)}.

\begin{figure}[h]
\begin{lstlisting}[language=Nickel]
let f = fun x y => x in
let g = f | (Num -> Num -> Num)
            @& (Num -> Bool -> Num) in
let h = g 1 in
(h 1, h true)
\end{lstlisting}
\caption{Intersection contracts don't distribute}
\label{fig:intersection-distribution}
\end{figure}

A solution, for the programmer, is to use an uncurried function
\nickel{fun (x, y) => x}. So one way to think about this limitation is
that currying function is not fully supported.

Note that if the last two lines of
Figure~\ref{fig:intersection-distribution} had read \nickel{(g 1 1, g
  1 true)} instead, then the coinductive semantics would accept the
example. It implies that under the coinductive semantics,
common-subexpression elimination (see Section~\ref{sec:optimizations})
is quite perilous.

\subsection{A First Realization}
\label{sec:keil-thiemann}

In Section~\ref{sec:issues-sem}, we've seen that different calls to a function
with a union contract must share information: the behavior of one call is
influenced by the previous ones, as the function must pick one component of the
union to satisfy across all usages. Conversely, following the semantics of
overloading, each application of a function with an intersection contract can
select a different branch and is thus independent from the others. A general
contract composed of nested unions, intersections and higher-order contracts
appears to require complex book-keeping in order to correctly raise
blame.

All of this still holds true of the coinductive semantics described in
Section~\ref{sec:coinductive-sem}.  Nevertheless,
\citeauthor{KeilThiemannUnionIntersection} give an algorithmic system which is
complete for their coinductive semantics. In a remarkable technical tour de
force, their algorithmic system allows for user-defined contracts (see
Section~\ref{sec:flat-and-inter}).

A key aspect of the approach of \citeauthor{KeilThiemannUnionIntersection}
is to rewrite nested union and intersection contracts into a
disjunctive normal form using the De Morgan's law \nickel{A @& (B @|
  C)}$\simeq$\nickel{(A @& B) @| (A @& C)}. The goal is to be able to
delay the choice of branch in intersections as much as possible.

To implement contract verification, \citeauthor{KeilThiemannUnionIntersection}
resort to specific reduction rules for unions and intersections which perform
this rewriting on the fly. This aspect is critiqued in
\citeauthor{RootCauseOfBlame}~\cite{RootCauseOfBlame}: ``the monitoring semantics for
contracts of intersection and union types given by Keil and Thiemann are not
uniform. (\ldots) If uniformity helps composition, then special cases can hinder
composition.''

A cost of this approach is that the De Morgan's law
\nickel{A @& (B @| C)}$\simeq$\nickel{(A @& B) @| (A @& C)} duplicates
contract \nickel{A}, which will cause some contracts to be checked
several times. This can be an issue with user-defined contracts which
may include costly tests.

Efficiency is also affected another way: each time a function with a
contract attached is applied, the whole \emph{context} must be
traversed to check for a compatibility property.

The algorithmic system of \citeauthor{KeilThiemannUnionIntersection}
is, on balance, a technically impressive realization of the coinductive
semantics that supports user-defined contracts, though it is fairly complex and
probably difficult to implement efficiently.

\subsection{Monitoring Properties}
\label{sec:will-morr-wadl}

Another realization of the coinductive semantics described in
Section~\ref{sec:coinductive-sem} is given in
\citeauthor{RootCauseOfBlame}~\cite{RootCauseOfBlame}, which aims at simplifying
the algorithmic system proposed by Keil and Thiemann and described in
Section~\ref{sec:keil-thiemann}.

A key ingredient of \citeauthor{RootCauseOfBlame} is to disallow user-defined contracts. This
choice gives the authors more freedom in the quest of a more uniform operational
semantics. This is sensible trade-off in the context of gradual typing à la
TypeScript: the problem is to match contracts with static types, and
user-defined contracts don't have a static type equivalent. On the other hand,
the cost would probably not be worth it for a configuration language like
Nickel.

%



As a means of proving the correctness of their simplified system, the
authors introduce what they call \emph{sound monitoring
  properties}. Here is the sound monitoring property for contexts of
intersection contracts:

$$ K \in \llbracket A \cap B \rrbracket^-~if~K \in \llbracket A \rrbracket^- \lor K \in \llbracket B \rrbracket^- $$

This reads as: a context $K$ satisfies the intersection of $A$ and $B$ if it
satisfies at least one of the two. Morally, the $K$s in $\llbracket A \cap B
\rrbracket^-$  should be the ones that can have their hole filled with a term
satisfying $A \cap B$ without violating the $A \cap B$ contract.

Although sound, this interpretation is weaker than what the
coinductive semantics permits. Consider the two contexts presented on
Figure~\ref{fig:valid-contexts}. The first one is a context satisfying
\nickel{Num -> Num}, applying the hole to a number. Similarly, the second context
from the same figure satisfies \nickel{Bool -> Bool}.

\begin{figure}[h]
\begin{lstlisting}[language=Nickel, title=\nickel{Num -> Num} context]
(*@$\square$@*) 3
\end{lstlisting}
\begin{lstlisting}[language=Nickel, title=\nickel{Bool -> Bool} context]
(*@$\square$@*) true
\end{lstlisting}
\caption{Two different contexts in Nickel}
\label{fig:valid-contexts}
\end{figure}

Now, combining these two contexts as in Figure~\ref{fig:invalid-context} gives a context
that doesn't satisfy \nickel{Num -> Num} nor \hbox{\nickel{Bool -> Bool}.}  According
to the sound monitoring property of intersection, Figure~\ref{fig:invalid-context} thus doesn't satisfy \nickel{Num
-> Num @& Bool -> Bool}.

\begin{figure}[h]
\begin{lstlisting}[language=Nickel]
let f = (*@$\square$@*) in
(f 3, f true)
\end{lstlisting}
\caption{Combined context}
\label{fig:invalid-context}
\end{figure}

The consequence is that \citeauthor{RootCauseOfBlame} don't prove their system
complete for the coinductive semantics.  It's probably just an oversight in the
proof: we believe their system to be indeed complete for the coinductive
semantics. But it does speak to the intrinsic complexity of union and
intersection contracts: it is remarkably easy to get details wrong.  This
difficulty contrasts with the standard framework of higher-order contracts where
satisfaction is much more straightforward.

\subsection{Racket}
\label{sec:racket}

Racket is a language based on the Scheme dialect of Lisp. Among established
languages, Racket is probably the one with the most comprehensive contract
system~\cite{RacketContracts}. Regarding union and intersection, Racket provides
the \racket{and/c} and \racket{or/c} combinators for contract.

The
\racket{and/c} combinator corresponds to the naive
interpretation of intersection described in
Section~\ref{sec:flat-and-inter}: applying contract \racket{(and/c A
  B)} is like applying contract \racket{A} then applying contract \racket{B}. In
particular \racket{and/c} doesn't model overloading. As in
Figure~\ref{fig:naive-impl}, the example given in
Figure~\ref{fig:racket-overloading} always fail because no argument satisfies
both \racket{number?} and \racket{string?}.

\begin{figure}[h]
\begin{lstlisting}[language=Racket]
(define/contract overload
(and/c (-> number? number?)
       (-> string? string?))
(lambda (x) x))
\end{lstlisting}
\caption{\racket{and/c} and overloading}
\label{fig:racket-overloading}
\end{figure}

The union combinator \racket{or/c} is similarly simple. It must be
able to decide immediately which branch holds: \racket{or/c} is a
simple Boolean disjunction. For instance, when higher-order contracts are combined using \racket{or/c}, Racket imposes
that contracts must be distinguishable by their arity. Doing so, there is at most
one candidate that can be selected directly. This is illustrated in
Figure~\ref{code:racket:or/c:working} whose program is accepted. On the other
hand, the program of Figure~\ref{code:racket:or/c:non-working} is rejected.

\begin{figure}[h]

\begin{lstlisting}[language=Racket]
(define/contract united
(or/c (-> number? number?)
      (-> string? string? string?))
(lambda (x) x))
\end{lstlisting}
\caption{Accepted use of \racket{or/c} with higher-order contracts}
\label{code:racket:or/c:working}

\end{figure}

\begin{figure}[h]

\begin{lstlisting}[language=lisp]
(define/contract united
(or/c (-> number? number?)
      (-> even? even?))
(lambda (x) x))
\end{lstlisting}
\caption{Rejected use of \racket{or/c} with higher-order contracts}
\label{code:racket:or/c:non-working}

\end{figure}

\paragraph{case->}

To compensate for the fact that \racket{and/c} doesn't support
overloading,
Racket provides a second intersection-like combinator: \racket{case->}. As for the \racket{or/c}, the candidate contracts must have
distinct arities to avoid ambiguity, unlike \racket{or/c} the
alternative chosen when the function is called rather than when the
contract is applied to the function. An example is provided in
Figure~\ref{fig:racket-case-fun}. The resulting possibilities are similar
to static overloading, where one function can take additional parameters for
example (e.g. as supported for Java methods). On the other hand, it excludes the
overloading of generic operations with fixed arity such as equality, comparison,
arithmetic operators, and so on.

\begin{figure}[h]
\begin{lstlisting}[language=Racket]
(define/contract overcase
  (case-> (-> string? string?)
         (-> number? number? number?)
  )
  (lambda (x [y 0]) (if (number? x)
                   (+ x y)
                   x)))

(overcase 1 2)

(overcase "hello")
\end{lstlisting}
\caption{Overloading with the \racket{case->} combinator}
\label{fig:racket-case-fun}
\end{figure}

In conclusion, Racket, a programming language with a large user base,
avoids the difficulties of general unions and intersections (Section~\ref{sec:issues-sem}). They make
the pragmatic choice of a simple semantics with limited support for
higher-order contracts in intersections and unions.

\section{(More) Related Work}
\label{sec:related-work}

\subsection{Higher-Order Contracts}

Enforcing pre- and post-conditions at runtime is a widely established practice.
In their foundational paper~\cite{FindlerFelleisenHOContracts}, Findler and
Felleisen introduce higher-order contracts, a principled approach to run-time
assertion checking that nicely supports functions. They introduce the notion of
blame, which is crucial to good error reporting. It became apparent later that
their contracts are closely related to the type casts introduced by gradual
typing, excluding blame: both \cite{FindlerMultiLang} and
\cite{FelleisenInterLang} see the value of contracts as a safe interface between
typed and untyped code. In \cite{WellTypedBlamed}, the authors precisely
introduce a system integrating gradual typing with contracts \textit{à la
Findler \& Felleisen}. Nickel adopts a similar type system, with both statically
typed terms, dynamically typed terms, and first-class contracts. Higher-order
contracts are the basis of the work of
\citeauthor{KeilThiemannUnionIntersection}~\cite{KeilThiemannUnionIntersection}
and \citeauthor{RootCauseOfBlame}~\cite{RootCauseOfBlame} that this paper
explored extensively.

\subsection{Unions and Intersections in Gradual Typing}

Castagna and Lanvin~\cite{CastagnaLanvinGradualUnionIntersection} introduce a
gradual type system based on set-theoretic
types~\cite{SetTheoreticTypes}.  Set-theoretic types feature unions,
intersections, negation types together with a notion of subtyping. This work
adheres to the \emph{static first school} (see~\cite{practiceTheoryGrad}) of
gradual typing: the reason for gradual types is to allow for a less precise type
information. It follows that their goals and constraints are somehow different,
resulting in the absence of first-class contracts. As for any gradual type
system, they do have to implement casts~---~that are very close to
contracts~---~for unions and intersections, but these casts are neither visible
nor available to the programmer. Obviously, no contracts mean no user-defined
contracts as well.

Castagna and Lanvin use abstract interpretation to derive the semantics of
unions and intersections. In their words: ``the resulting definitions are
quite technical and barely intuitive but they have the properties we seek for
[\ldots]''.  This makes any comparison with the coinductive semantics of
\citeauthor{KeilThiemannUnionIntersection} rather difficult. Castagna et
al.~\cite{gradualCastagna} build on Castagna and Lanvin to add polymorphism in
the picture. However, it is at the price of restricting the union and
intersection part: it is not possible to assign intersection types to a function
anymore.

\subsection{Context Sensitive Contracts}

The work by Dimoulas et al.~\cite{dimoulasESOP,dimoulasPOPL} presents
a complete study on systems that may execute (and potentially invalidate)
contracts inside other contracts, in particular through the study
of dependent function contracts.
Their ideas have the potential to help solve the problems outlined
on Sections \ref{sec:flat-and-inter} and \ref{sec:keil-thiemann},
however, it isn't clear how to leverage their techniques
for intersections.





\section{Conclusion}
\label{sec:conclusion}

Despite the fact that union and intersection of dynamic properties may
at first appear like an easy task, as soon as they are combined with
higher-order contracts for increased accuracy of error messages, they
really aren't.

The problem of union and intersection contracts is that they are not
orthogonal to apparently independent features of programming
languages. The mere presence of intersection and union contract
induces computational effects, and can make it quite difficult to
perform simple program optimizations such as inlining.

Designing a language with union and intersection contracts necessarily
means
making difficult choices: some features of union and intersection
contracts and of the rest of the language must be abandoned. Various
trade-offs can be made, but it is worth mentioning that implementing a
system with union and intersection contracts appears to be pretty
complex a task when unions and intersections are fairly complete.

It's hard not to have sympathy for the minimalist end of this
spectrum, where, like Racket, a language only has a very simple notion
of unions and intersections. Many applications of unions and
intersections are not possible in such a context, but the presence of
union and intersection contracts doesn't interact with the rest of the
language. It's probably more manageable, and it's the approach that
we are currently considering for Nickel.

To conclude, let us make clear that we do not think union and intersection
contracts are fundamentally broken, that they can not be implemented correctly,
or that they do not bear any value (quite the contrary). They may still make
sense to have in a language, and some apparent difficulties in the
implementation could be lifted some day. But as often, there are gaps between
the theoretical foundation, a proof-of-concept, a prototype, and the integration
in an actual language. We hope that our attempt may serve as a cautionary tale:
for union and intersection contracts, these gaps may be larger than
they appear.

\Urlmuskip=0mu plus 1mu
\printbibliography

\end{document}